\documentclass{ifacconf}

\usepackage{graphicx}      
\usepackage{natbib}        

\usepackage{url}
\usepackage{amsmath}
\usepackage{amsfonts}
\usepackage{graphicx}
\usepackage{amssymb}
\usepackage{alltt}
\usepackage{listings}
\usepackage[USenglish]{babel}
\usepackage{caption}
\usepackage{subfig}
\usepackage{enumerate}
\begin{document}
\begin{frontmatter}

\title{Integrated Control/Structure Design of a Large Space Structure using Structured $\mathcal{H}_\infty$ Control} 


\author[First]{J. Alvaro PEREZ} 
\author[Second]{Christelle PITTET} 
\author[Third]{Daniel Alazard} 
\author[Fourth]{Thomas Loquen} 

\address[First]{ONERA System Control Department, 
   Toulouse, 31000 France (e-mail: Jose-Alvaro.Perez\_ Gonzalez@onera.fr)}
   \address[Second]{CNES AOCS Department, 
   Toulouse, 31000 France (e-mail: christelle.pittet@cnes.fr)}
   \address[Third]{ISAE System Control Department, 
   Toulouse, 31000 France (e-mail: daniel.alazard@isae.fr)}
      \address[Fourth]{ONERA System Control Department, 
   Toulouse, 31000 France (e-mail: thomas.loquen@onera.fr)}
\begin{abstract}                
\quad This study presents the integrated control/structure design of a large flexible structure, the Extra Long Mast Observatory (ELMO). The integrated design is performed using structured $\mathcal{H_\infty}$ control tools, developing the Two-Input Two-Output Port (TITOP) model of the flexible multi-body structure and imposing integrated design specifications as $H_\infty$ constraints. The integrated control/structure design for ELMO consists of optimizing simultaneously its payload mass and control system for low-frequency perturbation rejection respecting bandwidth requirements.
\end{abstract}


\end{frontmatter}

\section{Introduction}
Currently Large Space Structures (LSS) are a challenging problem in control system design because they involve large complex kinematic chains composed of rigid and flexible bodies, mostly large-sized, maximally lightened, low-damped and with closed-spaced low natural frequencies. In this case structural modes interfere with the controlled bandwidth, provoking a critical Control-Structure Interaction (CSI). Therefore, LSS design is increasingly becoming subject to a close coordination among the different spacecraft sub-systems, demanding methods which tie together spacecraft structural dynamics, control laws and propulsion design. These methods are often called as \emph{Integrated Control/Structure Design} (ICSD), \emph{Plant-Controller Optimization} (PCO) or simply \emph{co-design} (CD).

ICSD methods began being studied in the 80s as an opposite technique to the current method of separated iterative sequences within the structural and control disciplines. The first integrated design methodologies were those in \cite{Haftka1987_ID}, \cite{Gilbert1988_ID} and \cite{Messac1992_ID}. These methods were based on iterative methodologies with optimization algorithms. Lately, other methods have been proposed such as those solved by LMI algorithms or with LQG methods like in \cite{Hiramoto2009_ID} and \cite{Cimellaro2008_ID} respectively. However, these approaches give conservative results and their applicability is restricted by problem dimension. Recently, a counterpart technique currently under development in ONERA Toulouse Research Center allows a more general approach \citep{Alazard2013_ID}.  Actually, this method is based on structured $\mathcal{H}_\infty$ synthesis algorithms developed in \cite{Gahinet2011_Hinf} or \cite{Burke2006_Hinf}, granting structured controllers and tunable parameters optimization. This synthesis, merged with a correct plant modeling, can reveal important applications of integrated design methodologies.

This work aims at showing the advances achieved at the end-way of this PhD study about an integrated design methodology with structured $\mathcal{H}_{\infty}$ control synthesis. This paper is organized as follows. First, the general framework about the integrated design used in this study is explained. This framework presents the modeling technique and optimization specifications that have been developed in order to be able to apply structured $\mathcal{H}_\infty$ synthesis. Second, ICSD is applied to a real case of a LSS: payload weight maximization and controller optimization for perturbation rejection are performed to a large flexible satellite composed of mast segments, developed by CNES space structures department. Finally, results of the ICSD are discussed.

\section{Integrated Design Method}
\label{sec:Method}

The ICSD method of this study lies on the structured robust control synthesis. A thorough explanation of structured $\mathcal{H}_\infty$ controller synthesis is given in \cite{Gahinet2011_Hinf} and \cite{Burke2006_Hinf}, where it is shown how is possible to impose the order, the structure and stability of the controller thanks to the structured $\mathcal{H}_\infty$ synthesis. In the following sections a descriptive view of the ICSD method is presented.

\subsection{Theory}
\label{subsec:hinf}
\begin{figure} 
\centering
\includegraphics[height = 5.5cm, width = 6cm]{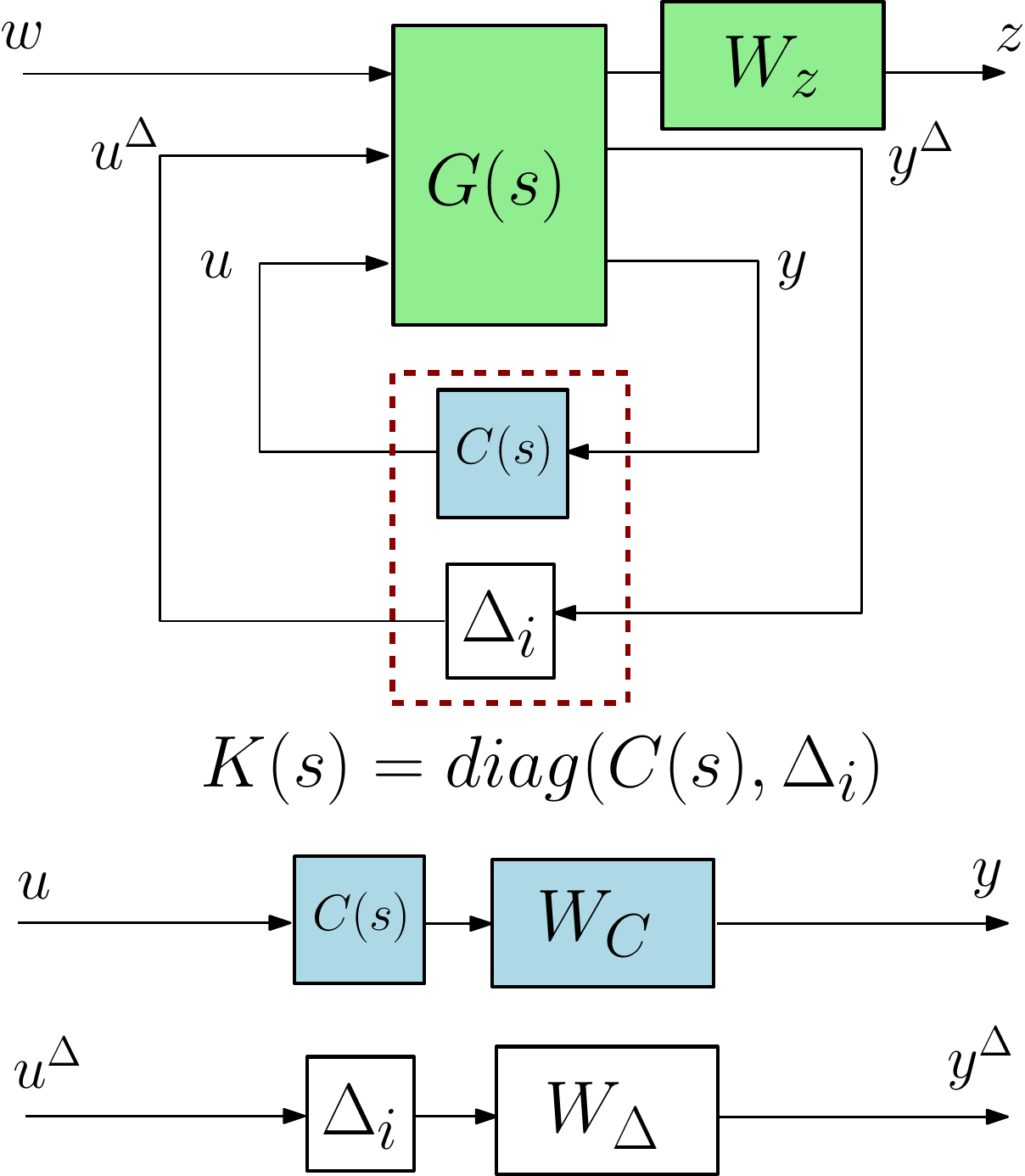}
\caption{Block Diagram of Integrated Design Optimization}
\label{fig:coDesignScheme}
\end{figure}

Figure \ref{fig:coDesignScheme} shows the standard multi-channel $\mathcal{H}_\infty$ synthesis problem. Given a Linear Fractional Representation (LFR) of the controlled system, $G(s)$, in which the corresponding parametric variations have been extracted as a tunable block $\Delta_i$, and added to an augmented structured controller with tunable parameters $K(s)= diag(C(s),\ldots,\Delta_i)$, structured $\mathcal{H}_\infty$ synthesis computes the optimal tuning of the free parameters of $C(s)$ and $\Delta_i$ to enforce closed-loop internal stability such that:

\begin{equation}
\label{eq:hinf}
a=a
\end{equation}

i.e., it minimizes the $\mathcal{H}_\infty$ norm between the perturbation input $w$ and the performance output $z$ such that it is constrained to be below $\gamma_{perf}>0$ to meet performances. The problem is in the form of \emph{Multi-Channel $\mathcal{H}_\infty$ synthesis}, and it allows imposing to the augmented controller different properties such as its internal stability \citep{Alazard2013_ID}, frequency template \citep{Loquen2012_Hinf} or maximum gain values. In substance, the \textbf{Structured $\mathcal{H}_\infty$ Integrated Design Synthesis} tunes the free parameters contained in the augmented controller $K(s) = diag(C(s),\Delta_i)$, $C(s)$ being a structured controller and $\Delta_i$ the set of parameters to be optimized, to ensure closed loop internal stability and meet normalized $\mathcal{H}_\infty$ requirements through $W_z$, $W_C$ and $W_\Delta$. Obviously, the difficulty lies on how to impose the correct normalized $\mathcal{H}_\infty$ requirements so that successful integrated design synthesis is guaranteed.   

\subsection{Modeling Technique}
\label{subsec:modelingTechnique}

As noted in Section \ref{subsec:hinf}, ICSD method with $\mathcal{H}_\infty$ control of a LSS needs a LFR representation of the different mechanical subsystems, so that parametric variations can be considered in the final plant model $G(s)$. A correct and straightforward modeling technique of a multi-body flexible spacecraft is the Two-Input Two-Output Port (TITOP) modeling technique \citep{Perez2015_LM, Alazard2015_LM, perez_IFAtheory2015, perez_IFAapply2015}, which allows casting different structural data (Finite Element Models, geometry, flexible modes) in the state-space domain, with the added possibility of including tunable variables. In substance, the different TITOP models of the different substructures can be easily assembled through load-acceleration transmission in order to reproduce the fully assembled LSS. 

This technique was born as a response to different control system needs. In preliminary design phases, the control engineer does not have an accurate description of the LSS to be controlled, only partial knowledge of the different substructures forming the LSS. In order to create a coherent fully assembled system on which control analysis can be performed, the TITOP modeling technique allows to easily connect the different substructures so that a coherent model of the system is obtained in LFR form. For example, Fig. \ref{fig:elmo} shows a LSS composed of a rigid hub, several flexible mast segments and a payload at its end. Currently the only information available is the one corresponding to one mast segment, without knowing the result of different mast segments linearly assembled. With the corresponding TITOP modelization, depicted in Fig. \ref{fig:elmoTITOP}, an accurate dynamic behavior of the assembled system is obtained in a straightforward manner.

\begin{figure}
\centering
\subfloat[]{\label{fig:substructure}\includegraphics[height = 3.2cm, width = 8.5cm]{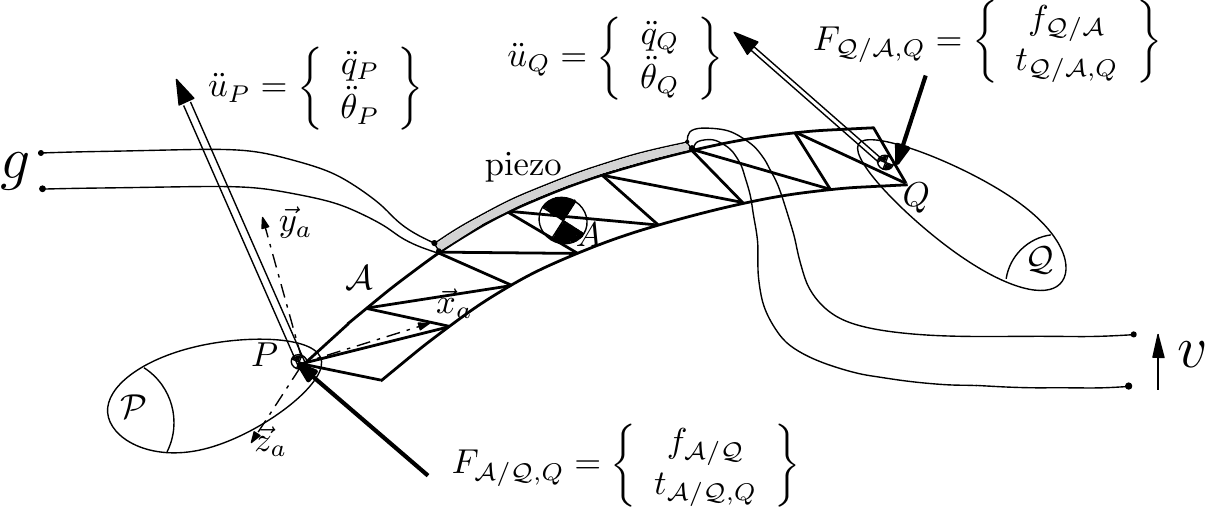}}
\hfill
\begin{minipage}{.48\linewidth}
\centering
\subfloat[]{\label{fig:titop} \includegraphics[height = 1.8cm, width = 4.2cm]{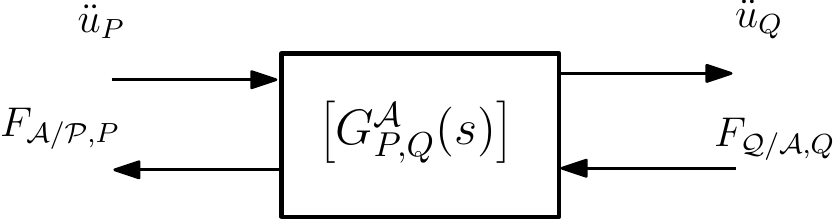}}
\end{minipage}%
\begin{minipage}{.48\linewidth}
\centering
\subfloat[]{\label{fig:pzttitop}\includegraphics[height = 1.8cm, width = 4.2cm]{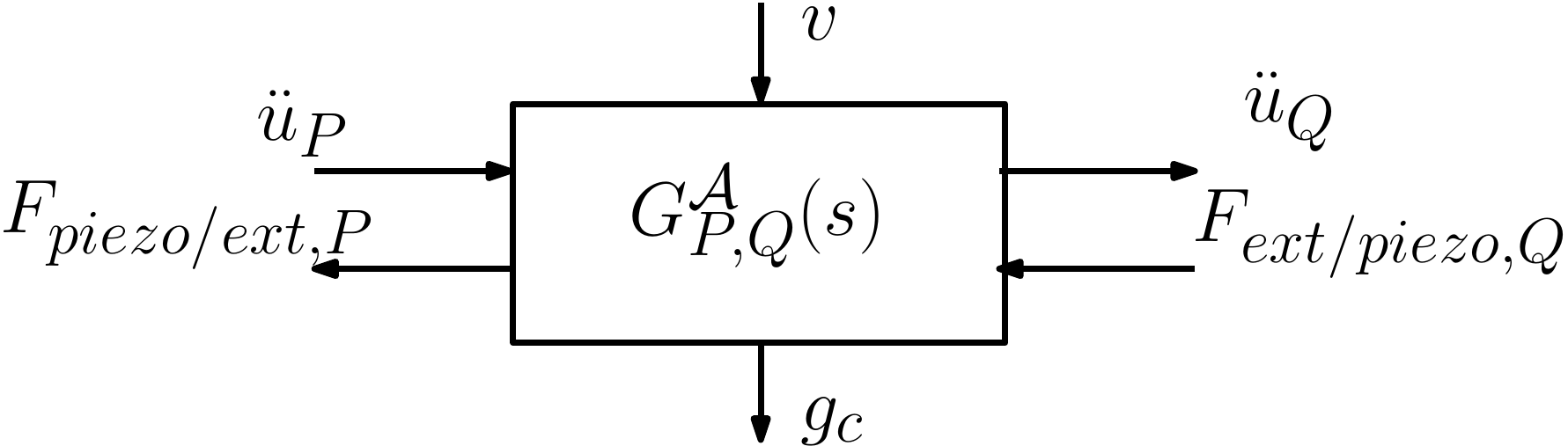}}
\end{minipage}
\caption{A generic substructure (a) in non-actuated TITOP form (b) and actuated TITOP form (c)}
\label{fig:main}
\end{figure}

TITOP models are \textit{ready-to-use} block diagrams in which a state-space representation of the substructure is cast. The state-space realization is obtained through application of the Component Modes Synthesis (CMS) \citep{Hurty1965_CM} and the Double-Port Approach (DP) \citep{Alazard2015_LM} to the equations of motion of the substructure, illustrated in Fig. \ref{fig:substructure}. There are two different types of TITOP models so far: actuated and non-actuated models.

\begin{itemize}
\item The \emph{non-actuated TITOP model} (see Fig. \ref{fig:titop}) of a substructure $\mathcal{A}$ represents a single mechanical subsystem between two mechanical subsystems, $\mathcal{P}$ and $\mathcal{Q}$, in which accelerations, $\ddot{u}$, are transmitted downstream, and loads, $F$, are transmitted upstream. Two connection points, $P$ and $Q$, are considered as the interfaces of the mechanical substructure with other substructures. The reader might head to \cite{Perez2015_LM} if more information about this modeling technique is desired.

\item The \emph{actuated TITOP model} (see Fig. \ref{fig:pzttitop}) has the same mechanical considerations as the \emph{non-actuated TITOP model}; i.e., has the same set of mechanical inputs/outputs. In addition, the model takes into account the electro-mechanical behavior of the substructure when piezoelectric elements are included inside the structure, adding additional inputs, the set of applied voltages $v$, and additional outputs, the set of measured electric charges $g_c$, which are suitable for accurate control action modeling. The reader might head to \cite{Perez2016_PEA} if more information about this extension of the TITOP modeling technique is desired.
\end{itemize}

Moreover, TITOP models can be enriched including parameter variations which can be used to create the final LFR model, $G(s)$ from which the set of tunable parameters $\Delta_i$ is extracted to be optimized.

\subsection{Integrated Design Specifications}
\label{subsec:IDspec}

The basic LSS design objectives for the control systems are (i) to obtain sufficiently high bandwidth and satisfactory closed-loop damping ratios for rigid-body structural modes; and (ii) to obtain satisfactory pointing errors. The first design requirement arises from the need to obtain a sufficient error decay when a disturbance occurs (such a sudden thermal distorsion when entering in Earth's shadow or a gravity gradient torque). The second design objective arises from mission performance requirements (such as alignement specification between two different on-board instruments, Radio Frequency specifications of a large antenna). These two objectives may not necessarily be compatible: increased feedback gains for obtaining higher bandwidth will in general lead to higher pointing errors since they may have an amplifying effect on sensor noise. 

In this method, control design objectives are adressed as frequency domain specifications on rigid-body modes and then expanded to the rest of the LSS where flexible motions have to be damped. The rigid modes are assigned to a specific part of the LSS, likely the hub, in which rigid-body actuators are usually placed to control attitude motion. The center of gravity of the hub is used as the origin of the frame for the rest of the constraints. 

The structured $\mathcal{H}_\infty$ synthesis scheme used in this method is based on the acceleration sensitivity function, $S_{\ddot{q}}$ \citep{Fezans2008_Hinf} i.e. the SOTAS (Second Order Template on Acceleration Sensitivity) scheme. This scheme measures the transfer between perturbations on acceleration vector $\ddot{q}$ and the performance output $z$, consisting of the same accelerations  weighted with a second-order filter:

\begin{equation}
\label{eq:hub2hub}
T_{w_{\ddot{q}}\rightarrow z} = W_{z} S_{\ddot{q}} = \frac{s^2 + 2 \xi_{req} \omega_{req}s + \omega_{req}^2 }{s^2} S_{\ddot{q}} 
\end{equation}

The weight function $W_{z}$ in Eq. (\ref{eq:hub2hub}) is a second order transfer function characterised by the desired closed-loop dynamics of the rigid-modes, to impose the desired frequency template to $S_{\ddot{q}}$. This ensures perturbation rejection at low frequencies in the desired bandwidth $\omega_{req}$ for the rigid-body modes, while high frequency disturbances (beyond controller's bandwidth) are not rejected. One of the main advantages of this synthesis scheme is that the optimal norm $\gamma_{opt}$ is equal to $1$ and therefore any additional constraint implies $\gamma_{perf} \geq 1$, being the distance to $1$ considered as a distance to the objective.

Once the specifications have been imposed to the rigid-body modes, the constraints can be expanded to other LSS locations subject to control objectives (such as antenna pointing or payload alignment with respect to the hub). These specifications are established through the transmission of the rigid-body motion to the other locations of the LSS subject to performance analysis. Therefore, the weight function for another location performance $z_P$ is stated as:

\begin{equation}
\label{eq:hub2tip}
T_{w_{\ddot{q}}\rightarrow z_P} = \phi_{GP} W_{z_G} S_{\ddot{q}} 
\end{equation}

where $z_G$ is the performance evaluation at the location where the rigi-body modes dynamics are considered (the hub), $G$, and $\phi_{GP}$ is the kinematic transport model from point $G$ to point $P$, where the constraint is imposed. This means that wherever the point $P$ is located in the LSS, its dynamics have to be the same as the rigid-body motion would induce in that location, minimising the effect of the flexible modes.

In the same manner, any disturbance affecting other locations in the LSS must be rejected at the hub following:

\begin{equation}
\label{eq:tip2hub}
T_{w_{\ddot{q}_P}\rightarrow z} = \phi_{PG} W_{z_P} S_{\ddot{q}_P} 
\end{equation}

Constraints for structural parameters $\Delta_i$ and controller $C(s)$ optimization can be imposed through the weightening filters $W_{\Delta_i}$ and $W_{C}$ respectively. Commonly $W_{C}$ is a roll-off specification to avoid the spill-over effect of neglected flexible modes. If parameter variation is correctly normalized in the interval $[-1,1]$ (value $0$ indicating the nominal value), a suitable weight function $W_{\Delta_i}$ can be a simple gain imposing the maximization or minimization of the different parameters. For example, if a parameter $\Delta_i$ has to be maximized, the constraint is expressed as:

\begin{equation}
\label{eq:parameter}
\min(\frac{z_{\Delta_i}}{w_{\Delta_i}})=\min(1-\Delta_i) \leq 1
\end{equation}

\section{Application to a LSS}

\subsection{Plant Description and Design Specifications}

\begin{figure} 
\centering
\includegraphics[height = 5cm, width = 7.5cm]{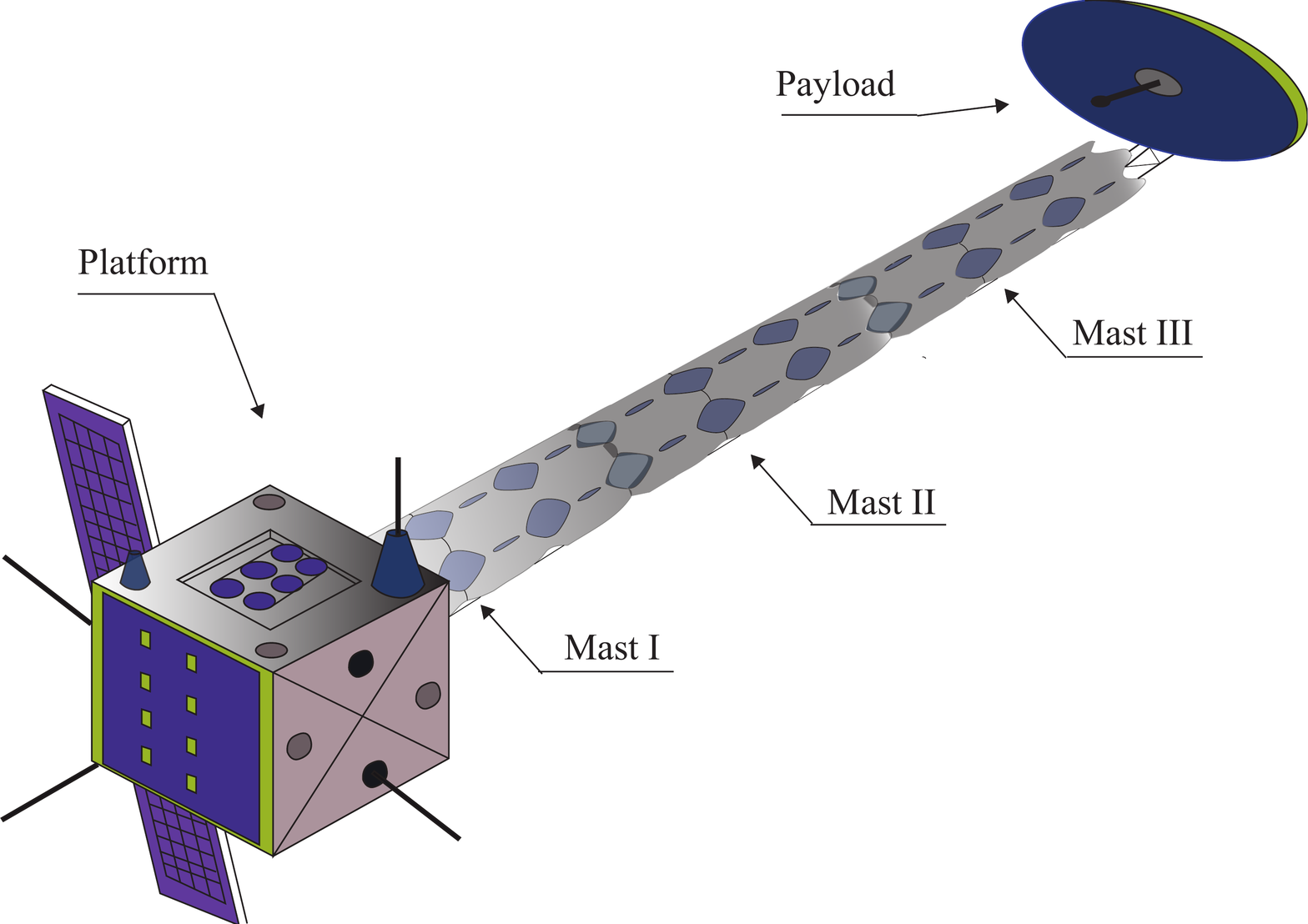}
\caption{Illustration of ELMO}
\label{fig:elmo}
\end{figure}

The Extra Long Mast Observatory (ELMO) study is composed of a rigid platform, with center of mass at $G$, to which a long deployable mast is cantilevered at a point $P$, composed of several mast segments, at which end the instrument for the required mission is located (depiction shown in  Fig. \ref{fig:elmo}). Therefore, ELMO will have a highly-directional inertia tensor, with a very high magnitude of the moment-of-inertia in the perpendicular plane to the mast. Thus the main challenge in science-mode control design is to provide just-enough bandwidth to adequately mitigate low-frequency disturbances while minimizing the effects of end displacement and maximizing instrument mass. ACS/ Structure design phase for ELMO is to provide valuable control strategies for the required bandwidth and to maximize the instrument mass located at the end of the mast.

The ELMO design specification requires rejection of low-frequency disturbances in a bandwidth estimated in the frequency range 0-0.173 Hz, mostly induced by the thermal-structural response of the system \citep{Park2002_SC}. In addition, the payload misalignment with the line of sight of the platform must not exceed 5 arcsec (0.08$^\circ$) during observation, which given the total length of the deployable mast can be translated to 17 mm of maximum tip displacement. The mass of the payload is expected to be around 60 Kg, but a maximization can be foreseen if this does not affect mission requirements.

\subsection{Plant Modeling and Assembly for ICSD}

The main difficulty of the ELMO study lies in the non-availability of a full plant model. Only data of a single mast segment is available, together with some platform and piezoelectric parameters referenced in Table \ref{tab:parameters}. To overcome this difficulty, the TITOP modeling technique is used to find a coherent model to proceed to ICSD synthesis.

\begin{figure} 
\centering
\includegraphics[height = 10cm, width = 7.5cm]{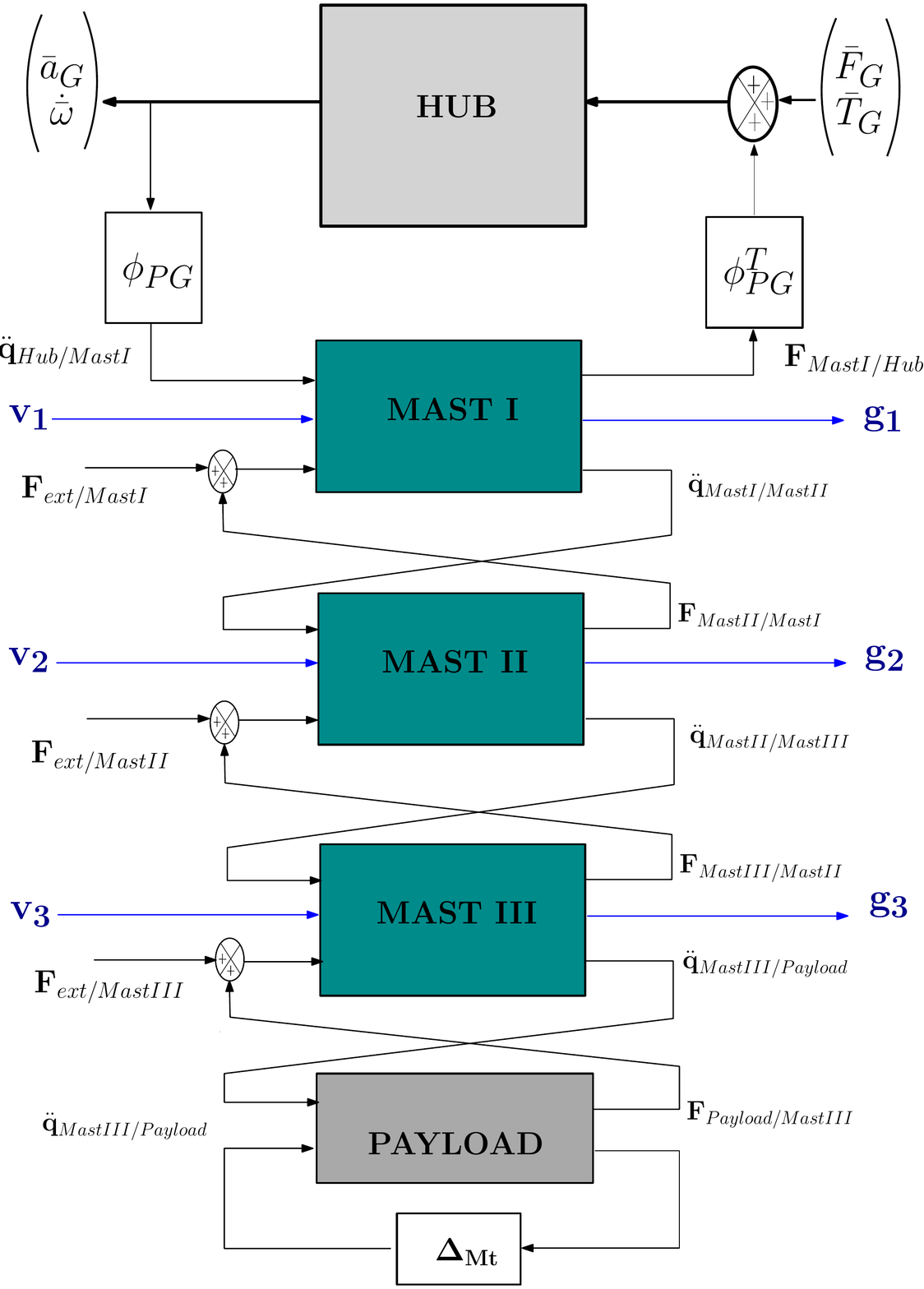}
\caption{TITOP modeling technique applied to ELMO study}
\label{fig:elmoTITOP}
\end{figure}

The assembly of the whole system is addressed as in Fig. \ref{fig:elmoTITOP}. Mast segments are connected as \textit{actuated TITOP} models, which exchange acceleration-loads through their connection points until the end of the mast, where the payload is placed. The masts can be actuated through the input of voltages $v_i$. The payload has been parametrized so that mass variation is taken into account through a tunable block $\Delta_m$ for maximization. The whole chain of flexible substructures is connected to the platform as a force feedback, previously transported by the transport dynamics matrix $\phi^T_{PG}$ which relates the connection point $P$ to the platform center of mass $G$.

Thus the TITOP modeling technique provides a coherent plant model from which ICSD can be performed. The inputs correspond to forces and torques applied to the platform together with the applied voltages in the piezoelectric actuators. The outputs correspond to the measured accelerations of the platform and payload misalignment, and displacement information extracted from the electric charge in the piezoelectric components. Final state-space assembly results in a first flexible bending mode at 0.62 Hz, close to the desired bandwidth, and a resulting inertia of 9800 Kg m$^2$.

\begin{table}
\caption{}
\begin{center}
\label{tab:parameters}
\begin{tabular}{c c c}
& & \\ 
\hline
\hline
Mast Parameters & Symbol & Value\\
\hline
Total length & $L$ & 4.060 m\\
Total weight& $m$ & 2.737 Kg\\
Average Thickness & $t$ & 9.53 mm \\
Diameter & $\phi$ & 141 mm \\
Average Elastic modulus & $E$ & 1.13 GPa \\
\hline
\hline
Actuator Parameters & Symbol & Value\\
\hline
Piezo length & $l_p$ & 80 mm\\
Thickness & $t_p$ & 2 mm \\
Width & $w_p$ & 30 mm \\
Volumetric Density & $\rho_p$ & 7600 Kg/m$^3$ \\
Elastic modulus & $E_p$ & 50 GPa \\
Piezoelectric Constant & $d_{31}$ & -150$\times 10^{-12}$ m/V \\
Dielectric Constant & $\epsilon_{33}^T$ & 1.59$\times 10^{-12}$ F/m \\
\hline
\hline
Hub Parameters & Symbol & Value\\
\hline
Hub inertia & $J_h$& 250 Kg m$^2$\\
Hub Mass & $m$& 500 Kg\\
Attachment Point & $P$& (0,1,0) m \\
\hline
\hline
Optimization Parameters & Symbol & Variation Range\\
\hline
Payload Mass & $Mt$ & $[60, 70]$ Kg \\
\hline
\hline
\end{tabular}
\end{center}
\end{table}

\subsection{Control Actions}

Taking advantage of the $\mathcal{H}_\infty$ synthesis, four control strategies are studied for perturbation rejection in order to compare their trade-off complexity-performance:

\begin{enumerate}[I]
\item One actuator located at the platform. Collocated PID controller for the control torque, $M_G$, applied by the reaction wheel. Platform's position $\theta$ is the controller input.
\begin{equation}
M_G^I = PID(s) \theta 
\end{equation}
\item One actuator located at the platform. Non-collocated PID -PD structured controller for the control torque applied by a reaction wheel at the platform. Platform's position $\theta$ and payload misalignment $y$ are the controller inputs.
\begin{equation}
M_G ^{II} = PID(s) \theta + PD(s) y
\end{equation}

\item One actuator located at the platform, piezoelectric actuators located at each mast segment. Collocated PID-PD structured controller for the torque and voltages control actions. Platform's position $\theta$ and electric charges $g_c$ at the piezos are the inputs of the controller.
\begin{equation}
\begin{aligned}
M_G ^{III} = PID(s) \theta \\
v_i ^{III} = PD(s) g_{c_i}
\end{aligned}
\end{equation}
\item One actuator located at the platform, piezoelectric actuators located at each mast segment. Non-collocated PID-PD-PI structured controller for torque and voltages control actions. Platform's position $\theta$, electric charges $g_c$ at the piezos and payload misalignment are the controller inputs.
\begin{equation}
\begin{aligned}
M_G ^{IV} = PID(s) \theta + PD(s) y \\
v_i ^{IV} = PD(s) g_{c_i} + PI(s) y
\end{aligned}
\end{equation}
\end{enumerate}

\subsection{Integrated Design Results}

The obtained ELMO model and the different controllers are implemented in the ICSD synthesis scheme of Fig. \ref{fig:coDesignScheme} and structured $\mathcal{H}_\infty$ synthesis is performed under the specifications presented in Section \ref{sec:Method}. The strategies are tested through perturbation rejection simulation for each co-designed system. The simulation scenario consists of a thermal torque induced by solar radiation, applied at the tip of the mast. Perturbation rejection and command actions are analyzed through time response plots. The analysis is done through 3 high-priority performance indexes: respect of the perturbation rejection template, $\gamma_{perf}$, hub pointing error $\delta \theta $ and payload misalignment $\delta y$. Low-priority indexes are the maximized payload mass for each strategy and the magnitude of the control commands for perturbation rejection (applied torque at the platform and applied voltages at the piezos).

\begin{figure*}[] 
  \begin{minipage}[b]{0.5\linewidth}
    \centering
    \includegraphics[width=8cm, height = 5cm]{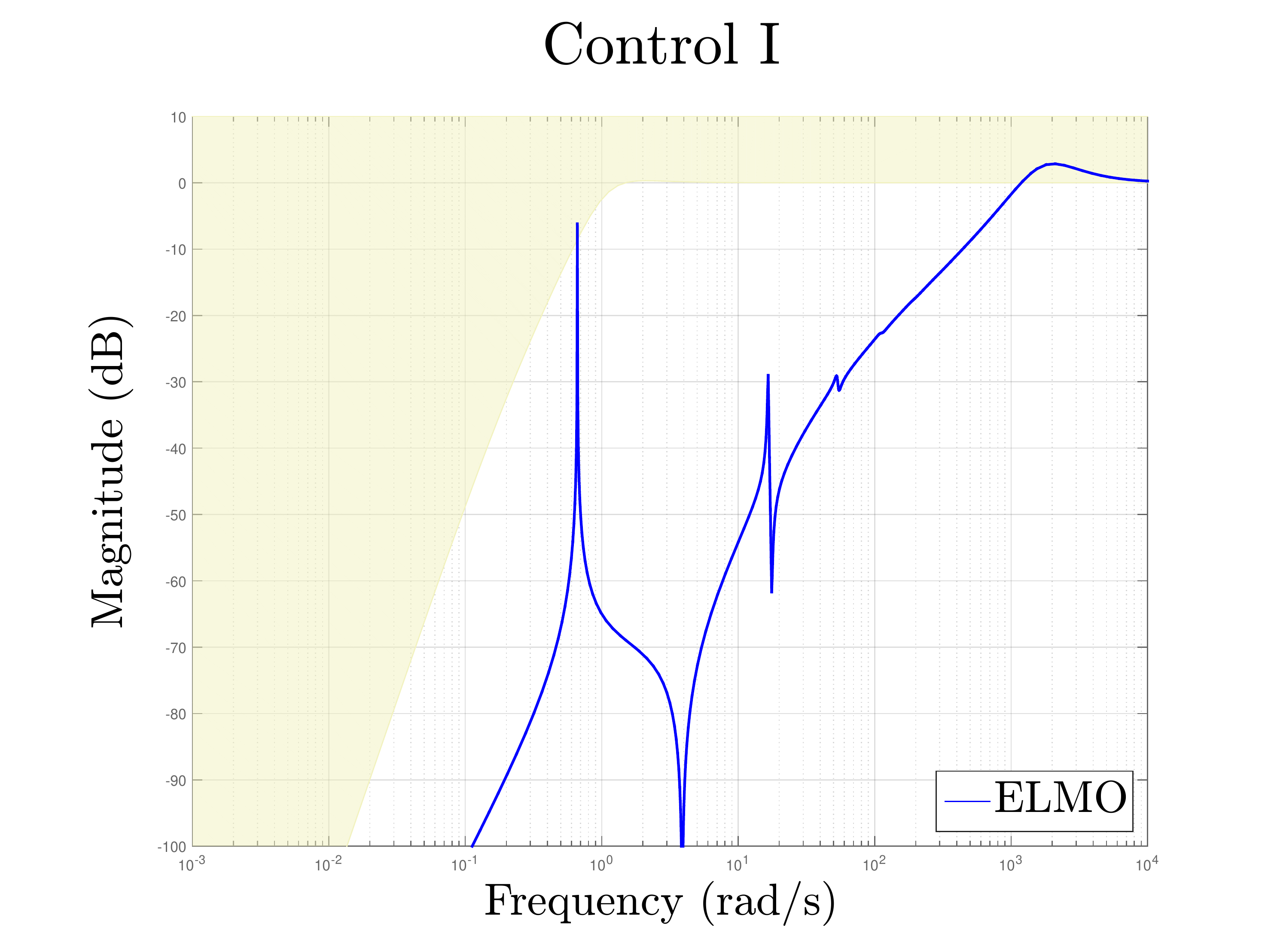}
     \label{fig:templateI}
  \end{minipage}
  \begin{minipage}[b]{0.5\linewidth}
    \centering
    \includegraphics[width=8cm, height = 5cm]{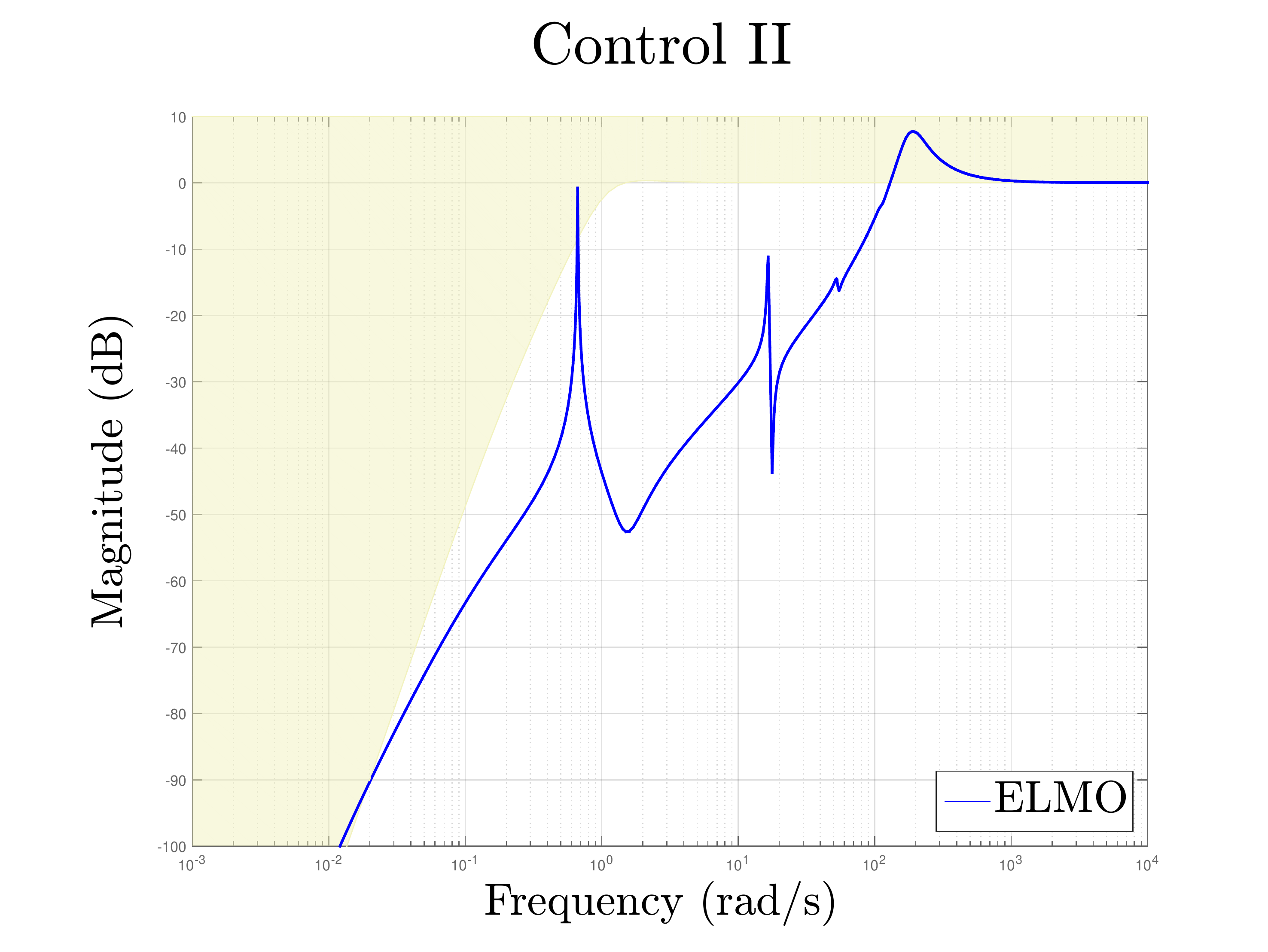} 
 	\label{fig:templateII}
  \end{minipage} 
  \begin{minipage}[b]{0.5\linewidth}
    \centering
    \includegraphics[width=8cm, height = 5cm]{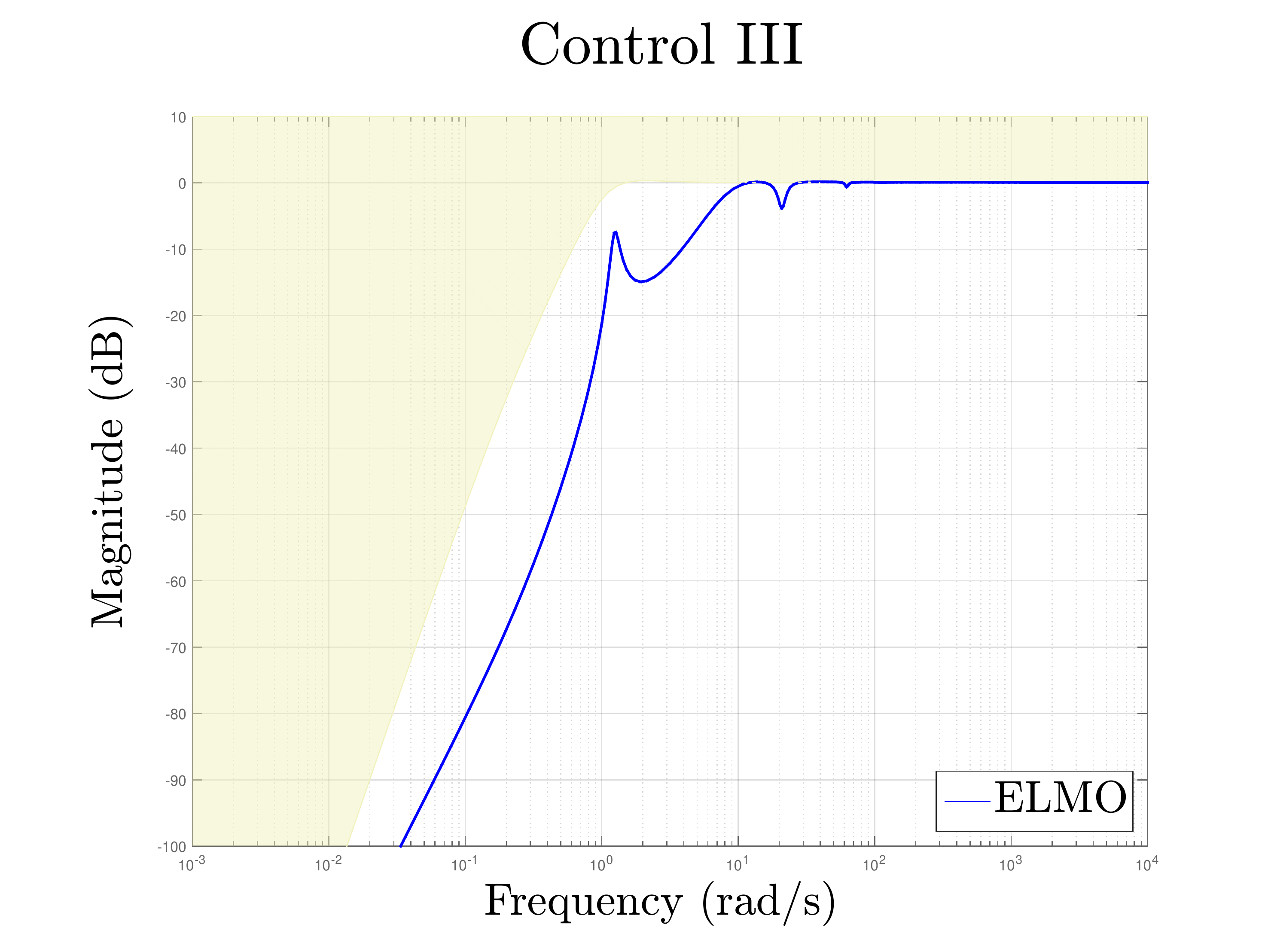} 
 \label{fig:templateIII}
  \end{minipage}
  \begin{minipage}[b]{0.5\linewidth}
    \centering
    \includegraphics[width=8cm, height = 5cm]{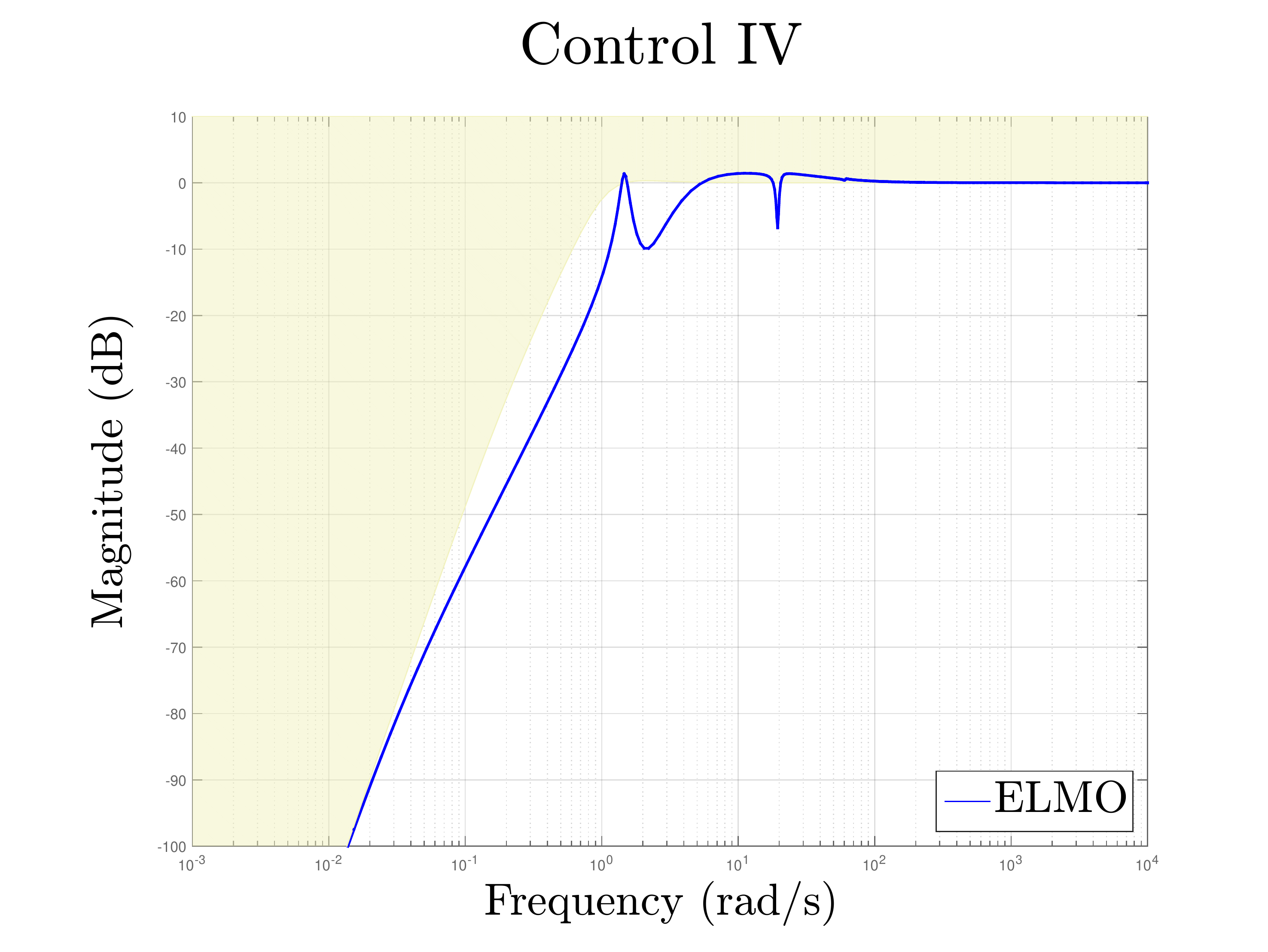} 
     \label{fig:templateIV}
  \end{minipage} 
  \caption{Acceleration Sensitivity Function Template (yellow background) and the obtained controlled system transfers}
  \label{fig:template} 
\end{figure*}

\begin{figure}[]
\centering
\subfloat{\label{fig:payloadMisalignment}\includegraphics[height = 5.5cm, width = 8.5cm]{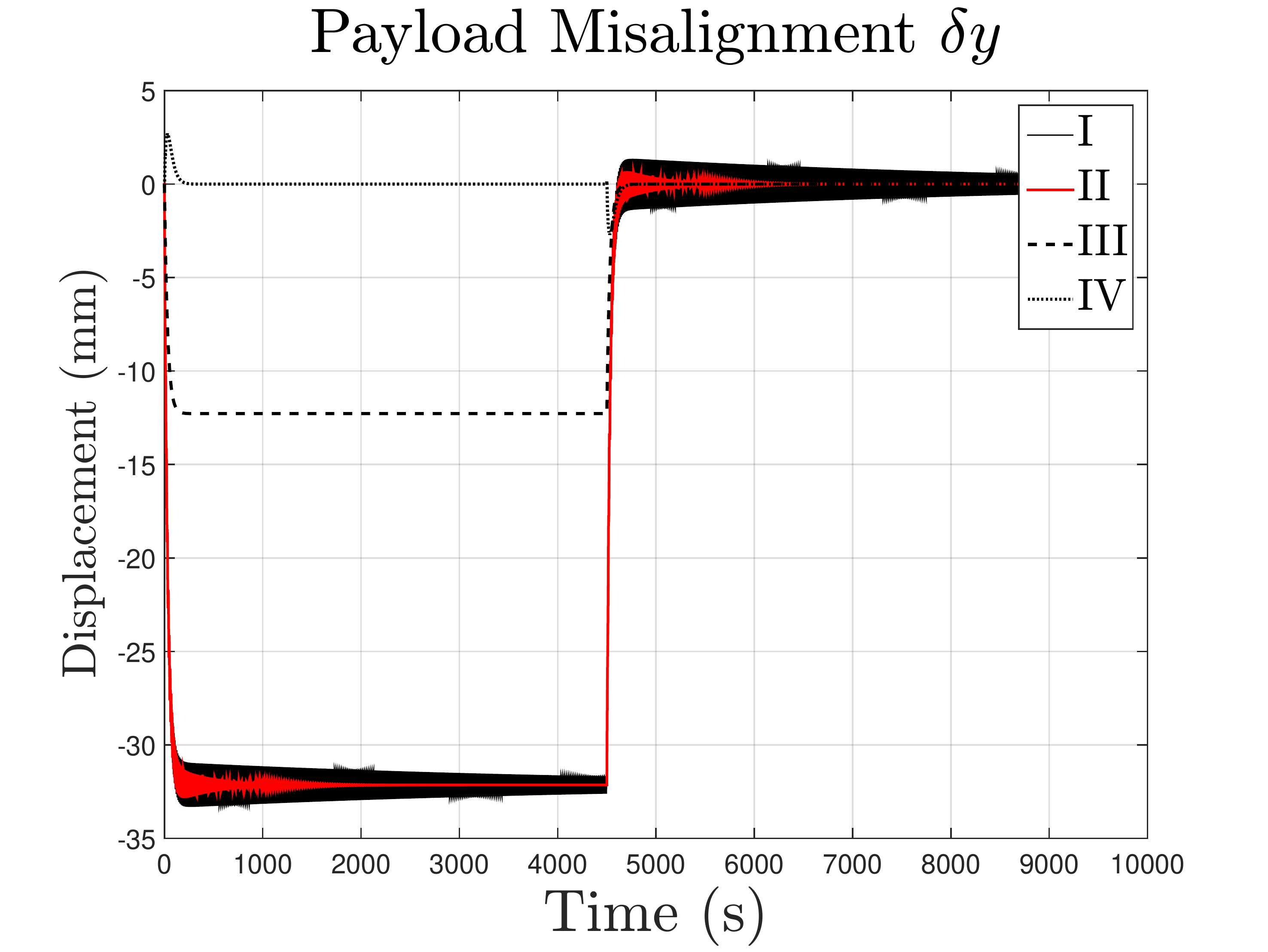}}
\hfill
\subfloat{\label{fig:hubPointing}\includegraphics[height = 5.5cm, width = 8.5cm]{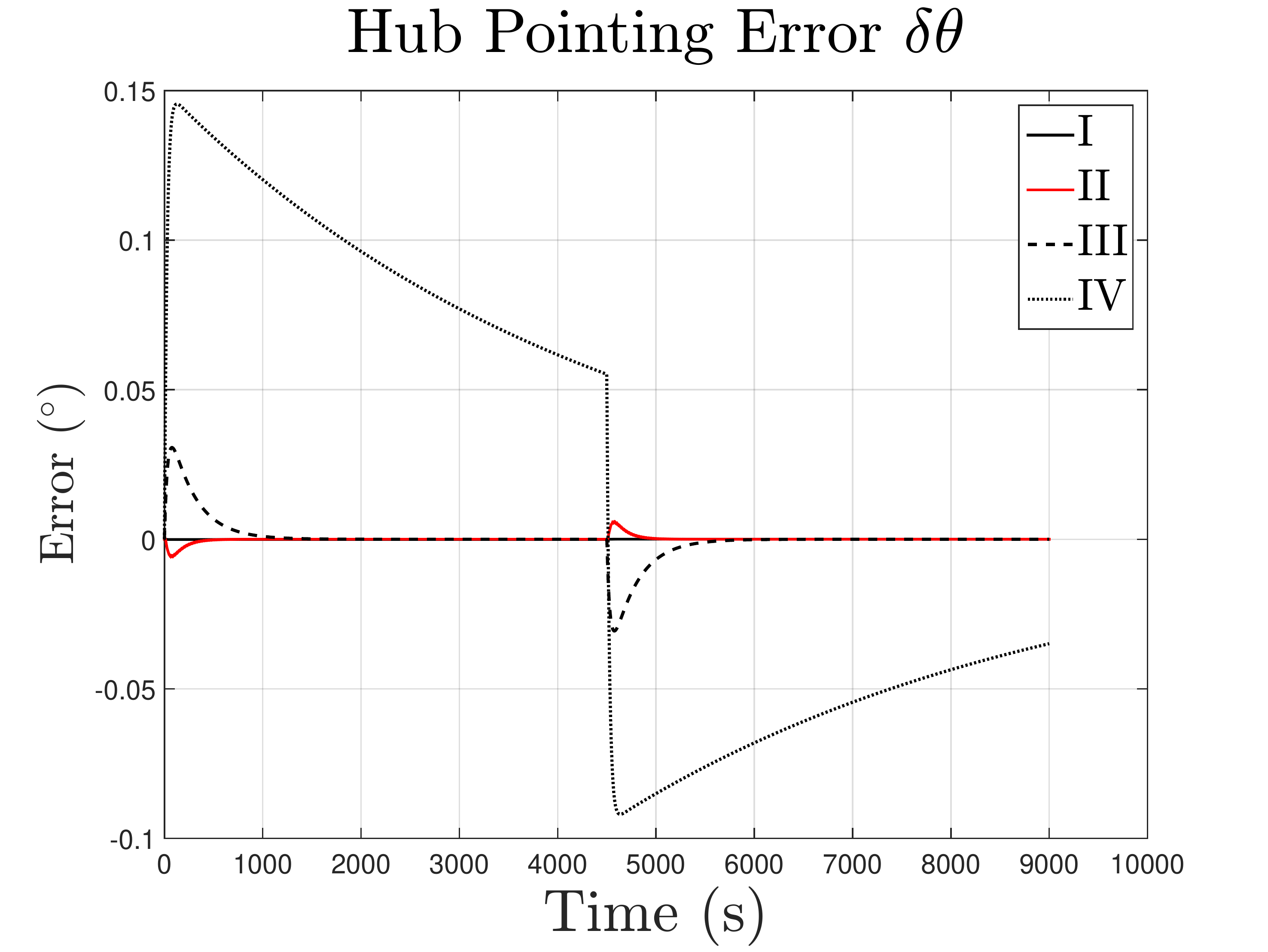}}
\caption{Response to thermal perturbation simulation: payload misalignment and hub pointing error}
\label{fig:simulationPointing}
\end{figure}

\begin{figure}[]
\centering
\subfloat{\label{fig:hubTorque}\includegraphics[height = 3cm, width = 7cm]{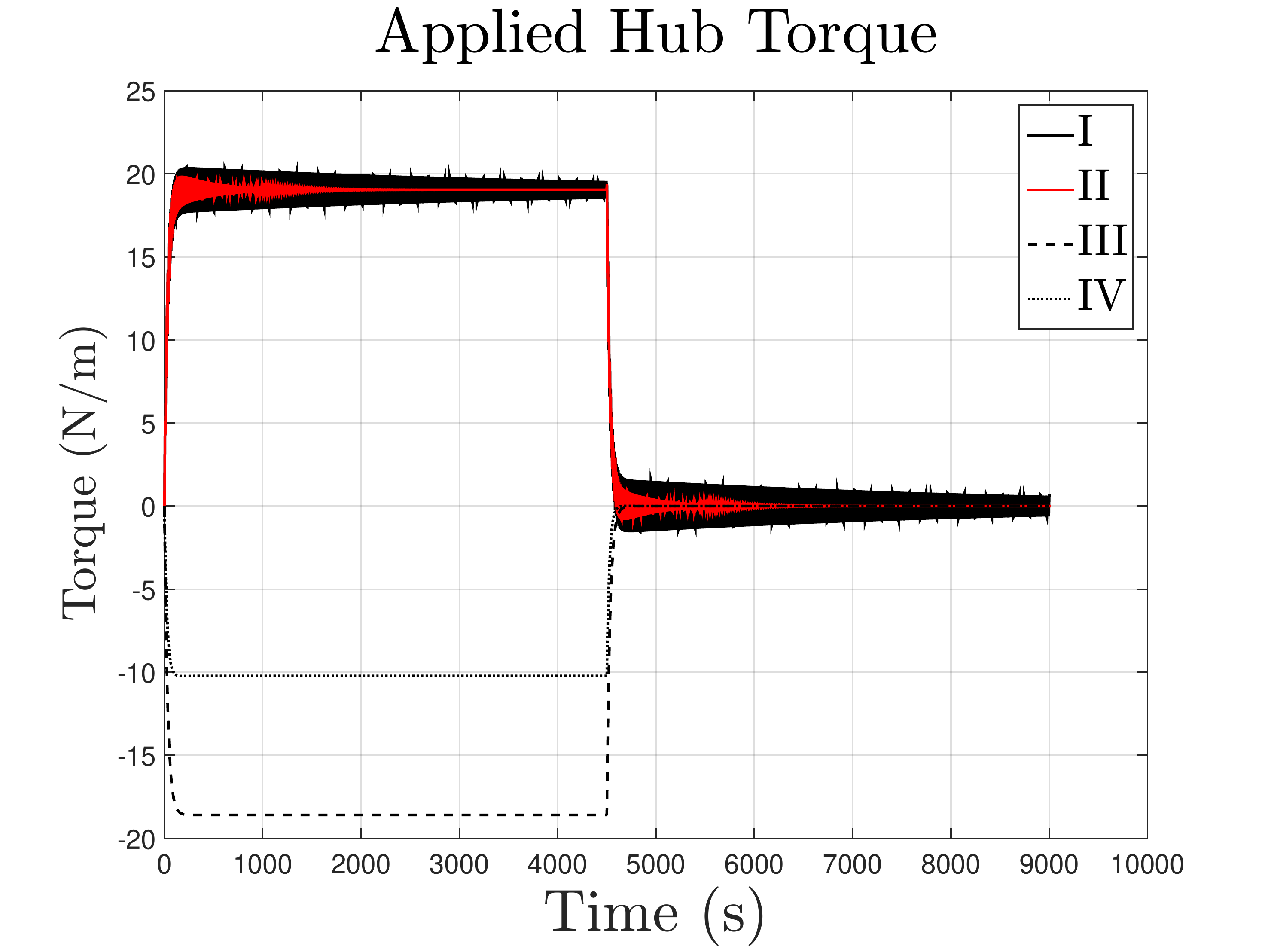}}
\hfill
\subfloat{\label{fig:voltageIII}\includegraphics[height = 3cm, width = 7cm]{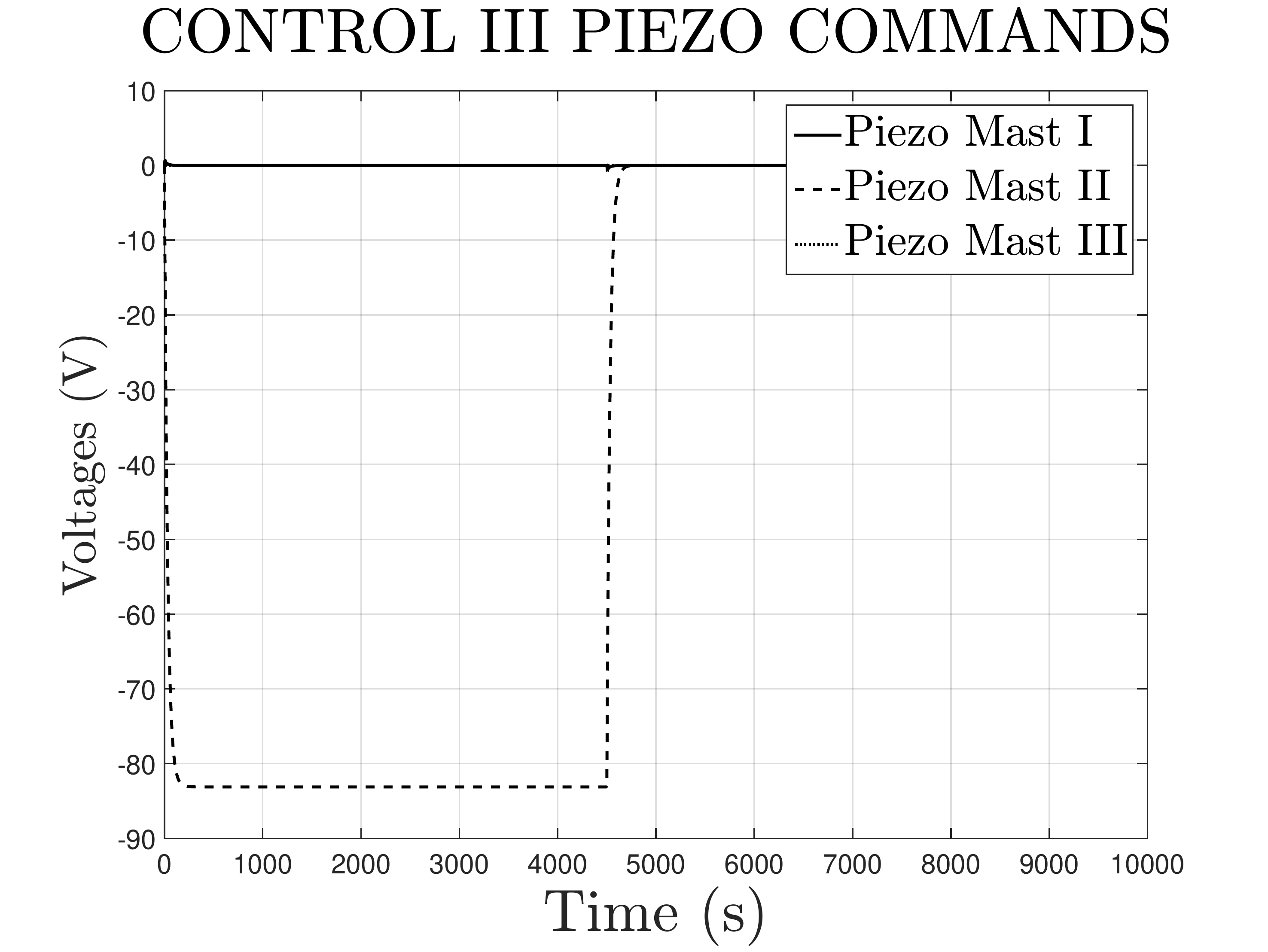}} 
\hfill
\subfloat{\label{fig:voltageIV}\includegraphics[height = 3cm, width = 7cm]{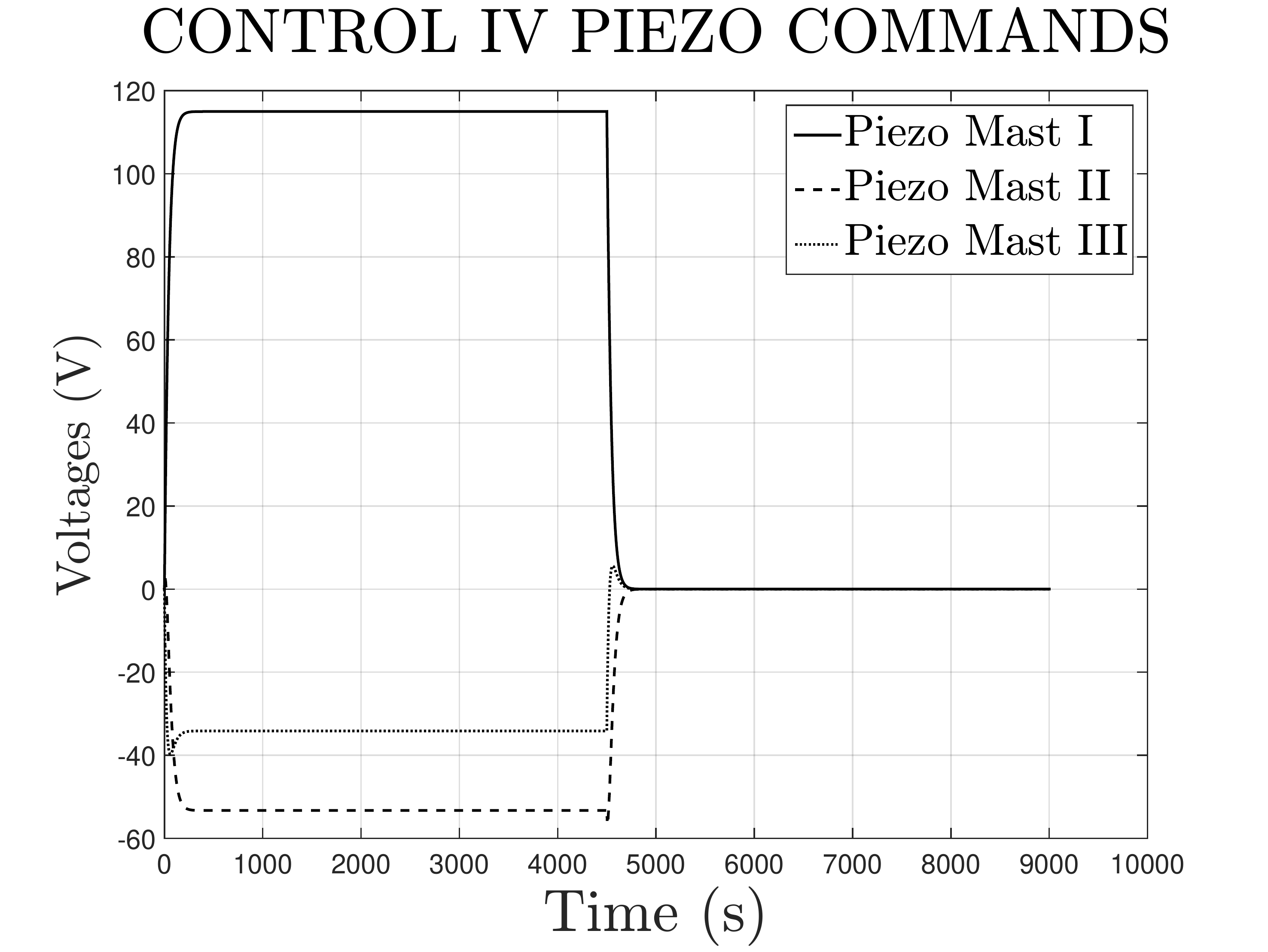}}
\caption{Response to thermal perturbation simulation: applied commands}
\label{fig:simulationCommands}
\end{figure}

Strategy I meets the perturbation requirement with $\gamma_{perf} = 1.3797$ (see Table \ref{tab:results}), respecting the frequency template until the first flexible mode (Fig. \ref{fig:templateI}) and with a fast hub pointing error rejection (Fig: \ref{fig:simulationPointing}). However, the payload misalignment is around 33 mm, exceeding the allowed 17 mm of maximum payload displacement. This is the consequence of the non-controllability of the bending mode by the platform's torque input. In addition, no attenuation vibration is possible, since the induced vibration frequency lies beyond the controller bandwidth, which requires active hub control torque even when the perturbation is not active (see Fig. \ref{fig:simulationCommands}). No payload maximization has been achieved with this strategy (see Table \ref{tab:results}).

\begin{table}
\caption{}
\begin{center}
\label{tab:results}
\begin{tabular}{c c c}
& & \\ 
\hline
\hline
Strategy & $\gamma_{perf}$ & Max Mt \\
\hline
I & 1.3797 &  60.82 Kg\\
II & 2.4231 & 60 Kg\\
III & 1.0088 & 63.94 Kg \\
IV & 1.1726 & 69.34 Kg \\
\hline
\hline
\end{tabular}
\end{center}
\end{table}

Strategy II is far from meeting the perturbation requirement with $\gamma_{perf} = 2.4231$ (see Table \ref{tab:results}), as it can be seen in Fig. \ref{fig:templateII}, where the sensitivity transfer function is comprised with the first flexible mode arising from the template. Hub pointing error rejection is achieved during simulation(Fig: \ref{fig:simulationPointing}), but payload exceeds maximum payload displacement. Induced vibrations are slightly more damped than those in Strategy I, but hub control torque is still required even when the perturbation is not active (see Fig. \ref{fig:simulationCommands}). No significant increase of payload mass has been achieved with this strategy.

Strategy III has the best peformance index for perturbation rejection, $\gamma_{perf} = 1.0088$, leading to the desired acceleration sensitivity template (see Fig. \ref{fig:templateIII}). Hub pointing error decay is slower than Strategies I and II, but always below the requirement of 5 arcsec. The static error of payload misalignment is  around 13 mm, below the maximum allowed. Vibrations are clearly damped during the simulation but the piezoelectric actuators are not equally required by the control system (piezo situated in mast II is over-used, Fig. \ref{fig:simulationCommands}). Around 4 Kg of extra payload mass are achieved during the optimization process. 

Strategy IV results in a performance index of $\gamma_{perf} = 1.1726$ penalized by the shadowing of the first flexible mode at the end of the bandwidth (see Fig. \ref{fig:templateIV}). Strategy IV has the slowest hub pointing error decay, exceeding the maximum allowed of 5 arcsec. On the contrary, payload misalignment presents the best behaviour with lack of static error misalignment thanks to the integral effect on tip displacement. Vibration damping is achieved and the payload mass maximization is around 10 Kg, the maximum allowed for this parameter.

The aforementioned results show that only Strategy III meets all the high-priority requirements: perturbation frequency template, payload misalignment and hub pointing error. In addition, it is possible to maximize payload mass by 3 Kg. However, applied commands are not realistic since only one piezoelectric component is used. Strategy IV could work if perturbation decay was faster, what can be done by further manual tuning of the integral effect. This is caused by the local-optimality of the structured $\mathcal{H}_\infty$ synthesis, which finds local optimums and the global optimality cannot be ensured. This is solved by random initialization of the tunable parameters \citep{Gahinet2011_Hinf} in the integrated design routine, but given the complexity of the problem, there are many solutions that can solve the optimization problem with different performances.

The feasibility of a structured $\mathcal{H}_\infty$ ICSD method has been proven by the ELMO study. Optimization constraints have been directly obtained from mission specifications and four possible control strategies have provided valuable answers for the control/structure design problem. In addition, the utility of using a general LSS modeling technique such as the TITOP modeling technique has been stated. However, further research must be done in order to reduce the difficulty of local optimality and improve other aspects of the control system such as minimum command energy or optimal actuator/sensor placement.

\section{Conclusions and Perspectives}

This study has presented the main guidelines for an Integrated Control/Structure Design Method using $\mathcal{H_\infty}$ synthesis and applied them to a Large Space Structure, ELMO, in order to study the expected performances with different control strategies. An intuitive and accurate modeling technique, the TITOP technique, has been explained and applied to a LSS, obtaining the plant model needed for integrated design. Frequency domain constraints have been established for structured $\mathcal{H}_\infty$ synthesis for integrated design, mostly focused on the perturbation rejection problem. Results have proven the advantages/disadvantages of the different control strategies, demonstrating the utility of ICSD in preliminary studies, with the added advantage of using a universal optimization tool such as structured $\mathcal{H}_\infty$ synthesis functions. 

Further applications of this methodology will involve optimal actuator placement, minimization of control power consumption through the introduction of $\mathcal{H}_2$ constraints in the integrated design scheme and developing new frequency templates for correct vibration attenuation directly obtained from design specifications. Further research will be done in order to reduce the difficulties associated to the non global optimality of the   structured $\mathcal{H}_\infty$ problem.

\begin{ack}
Research efforts were supported by the CNES AOCS Department and ONERA Flight Dynamics Department in Toulouse.
\end{ack}

\bibliography{mabiblioGENERAL}             
                                                   










\end{document}